# Deterministic Capacity of MIMO Relay Networks

Anders Høst-Madsen

**Abstract**

The deterministic capacity of a relay network is the capacity of a network when relays are restricted to transmitting *reliable* information, that is, (asymptotically) deterministic function of the source message. In this paper it is shown that the deterministic capacity of a number of MIMO relay networks can be found in the low power regime where $\mathrm{SNR} \to 0$. This is accomplished through deriving single letter upper bounds and finding the limit of these as $\mathrm{SNR} \to 0$. The advantage of this technique is that it overcomes the difficulty of finding optimum distributions for mutual information.

## I. Introduction

Recently there has been a large renewed interest in analyzing capacity of networks, in particular wireless networks. It has been found that the capacity of wireless networks can be increased by using the fact that the wireless signal propagates widely (the multicast advantage) and letting nodes cooperate (cooperative diversity) [1]. Coding methods for networks can generally be divided into two classes: those where relays process the received signal and forwards it, and those where the relays decode (a function of) the original message, and encodes this into a new signal. In the first class (some times denoted estimate-forward) are methods such as amplify-forward and compress-forward [2], [3] that have wide set of generalizations [??]. The second class (sometime denoted regenerative coding) has as its source the original decode-forward strategy of [4]. For a single antenna single relay channel [4]'s original strategy appears to be the only member of this class. However, for multiple antenna relay channels [5], [6] and multi-node networks [7], [8] there are many possible generalizations. What characterizes these methods, as opposed to the first class, is that relays decodes the original message, or more generally, a *function* of the original message, *reliably* (which also relates it to [9]), and transmits a message which is a possibly different function of the decoded message (e.g., in [8] the parity information). One way to characterize this class is that the transmission is reliable. Relays decode their messages with a vanishing error probability, and base their transmission on deterministic functions[1] of the messages. We will therefore denote this type of coding reliable coding, or deterministic coding. In contrast, amplify-forward type methods introduces further randomness through the noise at the relays. Informally one could say that amplify-forward type methods introduces errors in their transmission streams, while reliable transmission eliminates errors.

A. Høst-Madsen is with the Department of Electrical Engineering, University of Hawaii Manoa, Honolulu, HI 96822 (e-mail: madsen@spectra.eng.hawaii.edu. This work was supported in part by NSF grant CCF 0729152. This paper was presented in part at IEEE Information Theory Workshop, ITW 07, Lake Tahoe and Asilomar Conference on Signals and Systems, 2008.

[1]Deterministic in the sense that the functions do no depend on the noise realization in the network. Random encoding can be considered deterministic as nodes can share a common random number generator.

The aim of this paper is to bound the capacity that can be achieved with reliable transmission. Why reliable transmission? A number of practical arguments can be made in favor of reliable transmission: in large networks regeneration is needed at some point to avoid errors to accumulate. It is more similar to traditional multi-hop networking, making implementation potentially more smooth. It can seamlessly be combined with network coding [10]. When the the signal to noise ratio $\text{SNR} \to 0$ the noise in amplify-forward type becomes dominating, making the methods inefficient. The motivation for considering reliable transmission is this paper is mainly intellectual, though. It is a rather well characterized set of coding methods (think linear versus non-linear), and it is therefore of interests to find what rates can be achieved within this class of coding methods. Additionally, there are very few networks where the actual capacity can be found; even the simplest Gaussian relay channel has an unknown capacity. However, as will be seen, by restricting coding to this smaller class of methods, tight upper and lower bounds can be found for some networks under certain conditions. This is of course not the actual network capacity. However, it could be considered a restricted capacity, just as the capacity of a channel with modulation restricted to for example BPSK could be considered the BPSK capacity.

The rest of the paper is organized as follows. First we need a precise definition of what is meant by reliable transmission; this is provided in Section II. In Section III some results about the low power regime, $\text{SNR} \to 0$, are derived. As the main part, the capacity for reliable transmission of some MIMO relay networks is found in Section IV. Finally, generalizations of the results are discussed in section V.

## II. Definitions and Initial Remarks

We consider a network with $N$ nodes as in [11, Sec. 14.10]; each nodes may have multiple antennas. We denote the transmitted symbol (which might be a vector) at time $m$ at node $i$ by $x_i[m]$, the transmitted symbol from the $j$-th antenna by $x_{ij}[m]$, and the string of transmitted symbols in the interval $1 \dots k$ by $\underline{x}_{ij}[k] = [x_{ij}[1], x_{ij}[2], \dots, x_{ij}[k]]$. Similarly for the received signal $y_i[n]$ and $y_{ij}[n]$ at node $i$. A (length $n$) code for the network is defined as in [11, Sec. 14.10]: Node 1, the source has a message $W$ intended for node $N$ which it transmits at rate $R$; we consider the message a uniform random variable over $\{1, 2, \dots, 2^{nR}\}$. The encoder at node $i$ is a function (or code) $x_i[m](y_i[1 \dots m-1])$, $m \in \{1, \dots, n\}$ that depends only on *past* received symbols. The transmitted signal is a random variable $X_i[m]$ with the randomness coming from both the random message $W$ and from the noise in the received signal $y_i[m]$. The essence of deterministic capacity is to remove this latter randomness, with the following precise definition

*Definition 1:* A sequence of codes $\{\underline{x}_i[n], i = 1 \dots N\}$ is said to be *deterministic* or *reliable* if there exists a sequence of *deterministic* functions of the message $W$,$\{\underline{\hat{x}}_i[n](W), i = 1 \dots N\}$, so that

$$\forall i \in \{1, \dots, N\} \lim_{n \to \infty} P\{\underline{\hat{X}}_i[n] \neq \underline{X}_i[n]\} = 0 \tag{1}$$

$$\lim_{n \to \infty} P\{\hat{W}(\underline{Y}_N[n]) \neq W\} = 0 \tag{2}$$

The *deterministic capacity* is the supremum of all rates $R$ that is achievable by deterministic codes.

The definition is related to the computation rate in [9]. In this case, each node needs to compute a function of the message. This function is precisely the signal it is going to transmit. The principle is that there is no reason a node should decode more than needed for transmission. For example, if it transmits parity information about the message, this is all that it also needs to decode.

Equation (2) is the usual capacity condition of asymptotically zero decoding error probability. Equation (1) similarly states that, asymptotically, what node $i$ transmits depends only on the messages in the network, not the noise (realization). It therefore clearly excludes coding schemes such as amplify-forward [1] and compress-forward [4], [2], [3], but includes all decode-forward schemes known to the author: namely, in decode-forward a node decodes the message and forwards it. Since the condition is that this decoding happens with asymptotically zero error probability, it satisfies (1). However, the definition of deterministic capacity is much more general than specific decode-forward schemes, as is allows much flexibility of $\underline{\hat{x}}_i[n]$, including schemes such as those in [8].

One feature of definition 1 is that it allows usage of traditional methods of information theory. Equation (1) essentially says that node $i$ should be able to decode the function $\underline{\hat{X}}_i[n]$. One can then for example use Fano's inequality to outer bound the rate region. If there exist (decode-forward) coding methods achieving this outer bond, this is then the (deterministic) capacity.

While definition 1 applies to general channel models, We will in the following restrict attention to static, wireless channels with additive complex Gaussian noise of power $BN_0$, where $N_0$ is the noise power spectral density, and $B$ is the bandwidth. The static, complex, channel gain from node $i$ to node $j$ is $c_{ji}$, and if node $i$ has more than one antenna $c_{jik}$; we also define $\mathbf{c}_{ij} = [c_{ij1} \ldots c_{ijN}]$. We consider two case of channel state information (CSI):

- The synchronous case: all nodes are assumed to have full channel state information, i.e., to know perfectly all $c_{jik}$.
- The phase fading case[3], [12], [5]: all nodes know all $|c_{jik}|$, whereas the phase of $c_{ijk}$ is unknown to transmitters, but known at receivers. The phase of $c_{ijk}$ is assumed to vary ergodic during transmission. This can be used to model nodes that don't have synchronized local oscillators.

As a simple application of the definition we consider the one-relay relay network from [4], with a single relay and one antenna at all nodes. The received signals are

$$\underline{Y}_2[n] = c_{21}\underline{X}_1[n] + \underline{Z}_2[n] \tag{3}$$

$$\underline{Y}_3[n] = c_{31}\underline{X}_1[n] + c_{32}\underline{X}_2[n] + \underline{Z}_3[n] \tag{4}$$

Suppose $|c_{21}| \geq |c_{31}|$. Then node 2 can form

$$\underline{Y}'_3[n] = \frac{c_{31}}{c_{21}}\underline{Y}_2[n] + c_{32}\underline{X}_2[n] + \underline{Z}'_2[n] \tag{5}$$

$$= c_{31}\underline{X}_1[n] + c_{32}\underline{X}_2[n] + \frac{c_{31}}{c_{21}}\underline{Z}_2[n] + \underline{Z}'_2[n] \tag{6}$$

where $Z'_2[n]$ is iid Gaussian noise with power $1 - \frac{|c_{31}|^2}{|c_{21}|^2}$. Now consider the two companion signals

$$\underline{\hat{Y}}_3[n] = c_{31}\underline{X}_1[n] + c_{32}\underline{\hat{X}}_2[n] + \underline{Z}_3[n] \tag{7}$$

$$\underline{\hat{Y}}'_3[n] = \frac{c_{31}}{c_{21}}\underline{Y}_2[n] + c_{32}\underline{\hat{X}}_2[n] + \underline{Z}'_2[n] \tag{8}$$

By assumption node 3 can decode $W$ with small probability of error for large $n$. Since we consider deterministic capacity, we know that $\underline{Y}_3[n] = \underline{\hat{Y}}_3[n]$ with high probability for large $n$. A genie-aided node knowing $\underline{\hat{Y}}_3[n]$ therefore also can decode $W$ with small error probability (formally, the genie-aided node's error probability is bounded by $\hat{P}_e^{(n)} \leq P_e^{(n)} + P(\underline{Y}_3[n] \neq \underline{\hat{Y}}_3[n])$). Now, because $\underline{\hat{X}}_2[n]$ is a deterministic function of $W$, $\underline{\hat{Y}}_3[n]$ and $\underline{\hat{Y}}'_3[n]$ have the same distribution. Thus, a genie-aided node knowing $\underline{\hat{Y}}'_3[n]$ can also decode $W$ with small error probability. Finally, since $\underline{Y}'_3[n] = \underline{\hat{Y}}'_3[n]$ with high probability, a node knowing $\underline{Y}'_3[n]$ can also decode $W$ with small error probability. Thus, node 2 can decode $W$. A similar argument shows that for $|c_{21}| < |c_{31}|$ it does not help the destination to know $\underline{\hat{Y}}_3[n]$. Therefore, the rate is bounded by[2] $\max\{I(X_1; Y_2|X_2), I(X_1; Y_3|X_2)\}$, as well as the MAC bound $I(X_1, X_2; Y_3)$. On the other hand, Cover and El-Gamal's [4] block-Markov coding achieves

[2]The conditioning on $X_2$ enters the same way as the proof of Theorem 15.10.1 in [11]

this bound, and this *is* therefore the deterministic capacity. Essentially, this shows, not surprisingly, that Cover and El-Gamal's scheme is optimum among all decode-forward schemes.

It is difficult to extend the above example to larger networks. A major problem is proving that a Gaussian distribution is optimum. This problem can be overcome by working in the low power regime, and the rest of the paper will therefore restrict attention to this regime.

## III. The Low Power Regime

The capacity of the channel depends on the bandwidth as follows [11]: Fix $P$ (in Watts) and let the available bandwidth be $B$ (in Hz). The available power per sample is then $P/(2B)$ and the noise variance per sample $N_0/2$. If we denote by $C(B)$ the capacity (or spectral efficiency, [13]) in nats/s/Hz for a given bandwidth, we can define the following limit (if it exists)

$$\mathsf{C} = \lim_{B\to\infty} BC(B) \tag{9}$$

which is the limit of the capacity in nats/s for infinite bandwidth. We call the infinite bandwidth limit the low power regime; this has been considered in many papers, with the two papers [13], [14] breakthroughs. Signaling in the low power regime has a number of advantages: robustness to interference, little interference generation, covertness, etc., and is the principle behind UWB. For a point-to-point channel it is also the most energy efficient signaling. For multi-terminal channels it is not clear if this is still true, see e.g., [15].

The low power regime also has the theoretical advantage that $\mathsf{C}$ may be calculated without having explicit expressions for $C(B)$ using the techniques in [14] combined with the further results in [16], as we will see in the following.

We will denote rates in the low power regime by sans serif, i.e., if $\mathsf{R} \leq \mathsf{C}$ we say that the rate $\mathsf{R}$ (in nats/s) is achievable. Similarly, if $R \leq C(B)$, we say that the rate $R$ (in nats/s/Hz) is achievable.

We need the following generalizations of results in [16].

*Lemma 1:* Suppose that for each value of $B$ a random ($N$-vector) random variable $\mathbf{X}(B)$ that satisfy $\text{var}[\mathbf{X}(B)] \leq P$ is given[3]. Let $Y = \mathbf{c}^H\mathbf{X}(B) + Z$, where $Z \sim \mathcal{N}(0, N_0B)$. If $\mathbf{c}$ is a constant vector

$$\lim_{B\to\infty} BI(\mathbf{X}(B); Y) = \lim_{B\to\infty} \frac{\text{var}[\mathbf{c}^H\mathbf{X}(B)]}{N_0} \tag{10}$$

If $c_i = |c_i| e^{j\theta_i}$, where $\theta_i$ are iid random variables uniform on $[0, 2\pi]$ and $|c_i|$ constant then

$$\lim_{B\to\infty} BI(\mathbf{X}(B); Y|\boldsymbol{\theta}) = \lim_{B\to\infty} \frac{\sum_{i=1}^N |c_i|^2 \text{var}[X_i(B)]}{N_0} \tag{11}$$

*Proof:* The proof follows quite closely the proof of Lemma 1 in [16]. For completeness we will provide the proof in the asynchronous case. We can assume that $\mathbf{X}(B)$ has zero mean, as the mean will not influence the mutual information. Put $\tilde{Y} \sim \mathcal{N}\left(0, \sum_{i=1}^N |c_i|^2 \text{var}[X_i(B)] + N_0B\right)$, and write

$$I(\mathbf{X}(B); \mathbf{Y}|\boldsymbol{\theta}) = D\left(P_{\mathbf{Y}|\mathbf{X}(B)}||P_{\tilde{\mathbf{Y}}}|P_{\mathbf{X}(B)}, P_{\boldsymbol{\theta}}\right) \tag{12}$$

$$-D\left(P_{\mathbf{Y}}||P_{\tilde{\mathbf{Y}}}\,|\,P_{\boldsymbol{\theta}}\right) \tag{13}$$

[3] $\text{var}\mathbf{X}(B) = \text{tr}E\left[\mathbf{X}(B)\mathbf{X}(B)^H\right]$

The first term is

$$
\begin{aligned}
& D\left(P_{\mathbf{Y}|\mathbf{X}(B)}||P_{\tilde{\mathbf{Y}}}|P_{\mathbf{X}(B)},P_{\boldsymbol{\theta}}\right) \\
&= \int D\left(P_{\mathbf{Y}|\mathbf{X}=\mathbf{x}\boldsymbol{\theta}}||P_{\tilde{\mathbf{Y}}}\right)dP_{\mathbf{X}(B)}dP_{\boldsymbol{\theta}} && (14) \\
&= \int \log\left(\frac{\sum_{i=1}^{N}|c_i|^2\mathrm{var}[X_i(B)]+N_0B}{N_0B}\right)dP_{\mathbf{X}(B)}dP_{\boldsymbol{\theta}} \\
&\quad +\int \frac{\left|\mathbf{c}^H\mathbf{X}(B)\right|^2}{\sum_{i=1}^{N}|c_i|^2\mathrm{var}[X_i(B)]+N_0B} \\
&\quad +\frac{N_0B}{\sum_{i=1}^{N}|c_i|^2\mathrm{var}[X_i(B)]+N_0B}-1dP_{\mathbf{X}(B)}dP_{\boldsymbol{\theta}} && (15) \\
&= \log\left(1+\frac{\sum_{i=1}^{N}|c_i|^2\mathrm{var}[X_i(B)]]}{N_0B}\right) && (16)
\end{aligned}
$$

Since

$$
\begin{aligned}
\int\left|\mathbf{c}^H\mathbf{X}(B)\right|^2dP_{\mathbf{X}(B)}dP_{\boldsymbol{\theta}} &= \int \mathbf{c}^H E\left[\mathbf{X}(B)\mathbf{X}(B)^H\right]\mathbf{c}dP_{\boldsymbol{\theta}} && (17) \\
&= \sum_{i=1}^{N}|c_i|^2\mathrm{var}[X_i(B)] && (18)
\end{aligned}
$$

Thus

$$
\begin{aligned}
& \lim_{B\to\infty} BD\left(P_{\mathbf{Y}|\mathbf{X}(B)}||P_{\tilde{\mathbf{Y}}}|P_{\mathbf{X}(B)},P_{\boldsymbol{\theta}}\right) \\
&= \lim_{B\to\infty}\frac{\sum_{i=1}^{N}|c_i|^2\mathrm{var}[X_i(B)]}{N_0} && (19)
\end{aligned}
$$

The second term in (12) satisfies $\lim_{B\to\infty} BD\left(P_{\mathbf{Y}}||P_{\tilde{\mathbf{Y}}}|\,P_{\boldsymbol{\theta}}\right)=0$, which can be proven as follows. Fix $\boldsymbol{\theta}$. Then

$$\begin{aligned}
&\log\frac{P_{\mathbf{Y}}(y)}{P_{\tilde{\mathbf{Y}}}(y)}\\
&= \log\left(\frac{1}{\pi N_0 B}E\left[\exp\left(-\frac{1}{N_0B}\left|y-\mathbf{c}^H\mathbf{X}\right|^2\right)\right]\right)\\
&\quad -\log\left(\frac{1}{\pi\left(\sum_{i=1}^N |c_i|^2\mathrm{var}[X_i(B)]+N_0B\right)}\right.\\
&\quad \left.\exp\left(-\frac{1}{\sum_{i=1}^N |c_i|^2\mathrm{var}[X_i(B)]+N_0B}|y|^2\right)\right) \qquad (20)\\
&= \log\left(\frac{\sum_{i=1}^N |c_i|^2\mathrm{var}[X_i(B)]+N_0B}{N_0B}\right)\\
&\quad -\log E\left[\exp\left(\frac{1}{\sum_{i=1}^N |c_i|^2\mathrm{var}[X_i(B)]+N_0B}|y|^2\right.\right.\\
&\quad \left.\left.-\frac{1}{N_0B}\left|y-\mathbf{c}^H\mathbf{X}\right|^2\right)\right] \qquad (21)\\
&= \log\left(\frac{\sum_{i=1}^N |c_i|^2\mathrm{var}[X_i(B)]+N_0B}{N_0B}\right)\\
&\quad -\log E\left[\exp\left(\frac{1}{\sum_{i=1}^N |c_i|^2\mathrm{var}[X_i(B)]+N_0B}|y|^2\right.\right.\\
&\quad \left.\left.-\frac{1}{N_0B}\left(|y|^2-2\Re\left\{y\mathbf{c}^H\mathbf{X}\right\}+\left|\mathbf{c}^H\mathbf{X}\right|^2\right)\right)\right] \qquad (22)
\end{aligned}$$

Using series expansion we then get

$$\begin{aligned}
&\log\frac{P_{\mathbf{Y}}(y)}{P_{\tilde{\mathbf{Y}}}(y)}\\
&= \log\left(\frac{\sum_{i=1}^N |c_i|^2\mathrm{var}[X_i(B)]+N_0B}{N_0B}\right)\\
&\quad -\log E\left[1+o\left(\frac{1}{B}\right)+\frac{1}{N_0B}\left(2\Re\left\{y\mathbf{c}^H\mathbf{X}\right\}-\left|\mathbf{c}^H\mathbf{X}\right|^2\right)\right] \qquad (23)\\
&= \frac{\sum_{i=1}^N |c_i|^2\mathrm{var}[X_i(B)]}{N_0B}+o\left(\frac{1}{B}\right)\\
&\quad +o\left(\frac{1}{B}\right)+E\left[\frac{1}{N_0B}\left(2\Re\left\{y\mathbf{c}^H\mathbf{X}\right\}-\left|\mathbf{c}^H\mathbf{X}\right|^2\right)\right], \qquad (24)
\end{aligned}$$

where (24) uses the Lebesgue convergence Theorem to exchange limit and expectation. Then

$$\begin{aligned} E_{\boldsymbol{\theta}} \log \frac{P_{\mathbf{Y}}(y)}{P_{\tilde{\mathbf{Y}}}(y)} & \\ = & \frac{\sum_{i=1}^{N} |c_i|^2 \text{var}[X_i(B)]}{N_0 B} + o\left(\frac{1}{B}\right) \\ & + E_{\mathbf{X},\boldsymbol{\theta}}\left[\frac{1}{N_0 B}\left(2\Re\left\{y\mathbf{c}^H\mathbf{X}\right\} - \left|\mathbf{c}^H\mathbf{X}\right|^2\right)\right] \quad (25) \\ = & o\left(\frac{1}{B}\right) \quad (26) \end{aligned}$$

where we have used (18) and

$$\int \Re\left\{y\mathbf{c}^H\mathbf{X}\right\} dP_{\mathbf{X}(B)} dP_{\boldsymbol{\theta}} = 0 \quad (27)$$

■

*Lemma 2:* Suppose that for each value of $B$ we are given random variables $U(B)$, $V(B)$,and a random ($N$-vector) random variable $\mathbf{X}(B)$ that satisfy $\text{var}[\mathbf{X}(B)] \leq P$. Define

$$Y_1 = \mathbf{c}_1^H \mathbf{X}(B) + Z_1 \quad (28)$$

$$Y_2 = \mathbf{c}_2^H \mathbf{X}(B) + Z_2 \quad (29)$$

where $Z_1$ and $Z_2$ are independent, $Z_1, Z_2 \sim \mathcal{N}(0, N_0 B)$. Suppose that $(U(B), V(B)) \to \mathbf{X}(B) \to Y_1$ and $(U(B), V(B)) \to \mathbf{X}(B) \to Y_2$ form Markov chains. If $\mathbf{c}$ is a constant vector then

$$\begin{aligned} \lim_{B\to\infty} BI(U(B); Y_1) & \\ = & \lim_{B\to\infty} \frac{\text{var}[\mathbf{c}_1^H\mathbf{X}(B)]}{N_0} \\ & - \lim_{B\to\infty} \frac{\text{var}[\mathbf{c}_1^H\mathbf{X}(B)|U(B)]}{N_0} \quad (30) \\ \lim_{B\to\infty} BI(\mathbf{X}(B); Y_2|U(B)) & \\ = & \lim_{B\to\infty} \frac{\text{var}[\mathbf{c}_2^H\mathbf{X}(B)|U(B)]}{N_0} \quad (31) \\ \lim_{B\to\infty} BI(U(B); Y_1|V(B)) & \\ = & \lim_{B\to\infty} \frac{\text{var}[\mathbf{c}_1^H\mathbf{X}(B)|V(B)]}{N_0} \\ & - \lim_{B\to\infty} \frac{\text{var}[\mathbf{c}_1^H\mathbf{X}(B)|U(B), V(B)]}{N_0} \quad (32) \end{aligned}$$

If $c_{ij} = |c_{ij}|\, e^{j\theta_{ij}}$, where $\theta_{ij}$ are iid random variables uniform on $[0, 2\pi]$ and $|c_{ij}|$ constant then

$$\begin{aligned}
&\lim_{B\to\infty} BI(U(B); Y_1|\boldsymbol{\theta}) \\
&= \lim_{B\to\infty} \frac{\sum_{i=1}^{N} |c_{1i}|^2 \mathrm{var}[X_i(B)]}{N_0} \\
&\quad - \lim_{B\to\infty} \frac{\sum_{i=1}^{N} |c_{1i}|^2 \mathrm{var}[X_i(B)|U(B)]}{N_0} \qquad (33)\\
&\lim_{B\to\infty} BI(\mathbf{X}(B); Y_2|U(B), \boldsymbol{\theta}) \\
&= \lim_{B\to\infty} \frac{\sum_{i=1}^{N} |c_{2i}|^2 \mathrm{var}[X_i(B)|U(B)]}{N_0} \qquad (34)\\
&\lim_{B\to\infty} BI(U(B); Y_1|V(B), \boldsymbol{\theta}) \\
&= \lim_{B\to\infty} \frac{\sum_{i=1}^{N} |c_{1i}|^2 \mathrm{var}[X_i(B)|V(B)]}{N_0} \\
&\quad - \lim_{B\to\infty} \frac{\sum_{i=1}^{N} |c_{1i}|^2 \mathrm{var}[X_i(B)|U(B), V(B)]}{N_0} \qquad (35)
\end{aligned}$$

Assuming all limits are defined.

*Proof:* For (31,34), we use a conditional version of Lemma 1. For (30,33) we write $I(U(B); Y_1) = I(\mathbf{X}(B); Y_1) - I(\mathbf{X}(B); Y_1|U(B))$ (using the Markov chain property). For (32,35) we write $I(U(B); Y_1|V(B)) = I(\mathbf{X}(B); Y_1|V(B)) - I(\mathbf{X}(B); Y_1|U(B), V(B))$. ■

## IV. Deterministic Capacity of MIMO relay channels

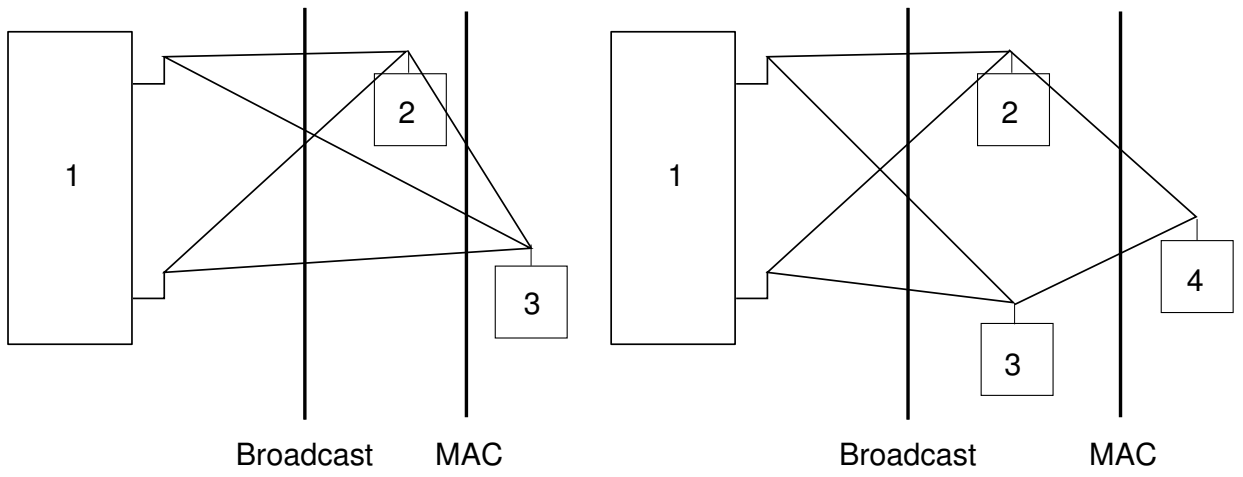

Fig. 1. The relay channels considered.

In this section we will find the low power deterministic capacity of the channels in Fig. 1 under certain conditions. Define

$$W_i = \underline{\hat{X}}_i[n] \qquad (36)$$

and notice that for deterministic capacity, node $i$ must be able to decode $W_i$. Specifically, we have the following statement of Fano's inequality, proven similarly to Fano's inequality in [11].

*Lemma 3 (Fano's inequality):* Suppose that the source message $W \in \left\{1 \ldots 2^{nR}\right\}$. If a node uses the deterministic sequence of codes $\underline{X}_i[n]$ the following inequality holds

$$1 + Pr\left\{\underline{X}_i[n] \neq \underline{\hat{X}}_i[n]\right\} nR \geq H\left(\underline{\hat{X}}_i[n] | \underline{Y}_i[n]\right) \qquad (37)$$

Now define for $i,j \in \{2,3\}$

$$R_i = \lim_{n\to\infty} \frac{H(W_i)}{n} \tag{38}$$

$$R_{ij} = \lim_{n\to\infty} \frac{H(W_i|W_j)}{n} \tag{39}$$

$$R = \lim_{n\to\infty} \frac{H(W_i,W_j)}{n} \tag{40}$$

We of course have $R = R_{23} + R_3 = R_{32} + R_2$.

The channel from node 1 to (2,3) is a MIMO broadcast channel in both networks. The MIMO broadcast channel was considered in [17] and [18]. However, in our case arbitrary dependency between the messages are allowed, and we therefore cannot directly use the results of [18]. However, we can prove the following *outer bound*, a generalization (to dependent messages) of Lemma 3.5 in [19]

*Proposition 1:* The capacity region of the broadcast channel is contained in the convex closure of all $(R, R_2, R_3)$ that satisfy

$$R_2 \leq I(U_2;Y_2) \tag{41}$$

$$R_3 \leq I(U_3;Y_3) \tag{42}$$

$$R = R_{23} + R_3 \leq I(U_2;Y_2|U_3) + I(U_3;Y_3) \tag{43}$$

$$R = R_{32} + R_2 \leq I(U_3;Y_3|U_2) + I(U_3;Y_3) \tag{44}$$

for some joint distribution $p(u_2,u_3)p(x|u_2,u_3)$, where $\mathrm{var}[\mathbf{X}] \leq P_1$.
The proof follows quite closely that of [19], so we will not provide it here.

*Theorem 1:* The deterministic capacity of the relay channel in Fig. 1(a) in the low power regime in the synchronous case is given by maximizing

$$\begin{aligned}\mathsf{R} \leq\ & (|c_{311}|^2 + |c_{312}|^2)P_{31} \\ & + (|c_{211}|^2 + |c_{212}|^2)\cos^2(\alpha-\theta)P_{21}\end{aligned} \tag{45}$$

$$\begin{aligned}\mathsf{R} \leq\ & (|c_{311}|^2 + |c_{312}|^2)P_{31} \\ & + (|c_{311}|^2 + |c_{312}|^2)\cos^2(\theta)P_{21} \\ & + \left(\sqrt{P_{b1}(|c_{311}|^2 + |c_{312}|^2)} + |c_{32}|\sqrt{P_2}\right)^2\end{aligned} \tag{46}$$

with respect to $P_{21}$, $P_{31}$, $P_{b1}$, and $\theta$, subject to $P_{21} + P_{31} + P_{b1} \leq P_1$. Here $\alpha = \arccos\left(\frac{|\mathbf{c}_2^H\mathbf{c}_3|}{\|\mathbf{c}_2\|\|\mathbf{c}_3\|}\right)$. In the phase fading case, the capacity is given by

$$\begin{aligned}\mathsf{C} = \ & \min\big\{\max\{|c_{311}|^2 + |c_{312}|^2, \\ & |c_{211}|^2 + |c_{212}|^2\}P_1, \\ & (|c_{311}|^2 + |c_{312}|^2)P_1 + |c_{32}|^2P_2\big\}\end{aligned} \tag{47}$$

*Proof:* The rate is bounded by

$$R \leq I(X_1;Y_3|U_2,X_2) + I(U_2;Y_2|X_2) \tag{48}$$

$$R \leq I(X_1,X_2;Y_3) \tag{49}$$

The bound (49) is simply the MAC bound into the destination. In the following use the notation $\overline{X}[m] =$

$[X[m], X[m+1], \ldots, X[n]]$. The bound (48) is then proven by the following chain of inequalities

$$
\begin{aligned}
&nR \\
&= H(W) \; = \; H(W_2) + H(W|W_2) && (50) \\
&= H(W_2|\underline{Y}_2[n]) + I(W_2;\underline{Y}_2[n]) \\
&\quad +H(W|\underline{Y}_3[n], W_2) + I(W;\underline{Y}_3[n]|W_2) && (51) \\
&\leq n\epsilon_n + I(W_2;\underline{Y}_2[n]) + I(W;\underline{Y}_3[n]|W_2) && (52) \\
&= n\epsilon_n + \sum_{m=1}^{n} I(W_2;Y_2[m]|\underline{Y}_2[m-1]) \\
&\quad + \sum_{m=1}^{n} I(W;Y_3[m]|W_2,\overline{Y}_3[m+1]) && (53) \\
&= n\epsilon_n + \sum_{m=1}^{n} H(Y_2[m]|\underline{Y}_2[m-1]) \\
&\quad - \sum_{m=1}^{n} H(Y_2[m]|W_2,\underline{Y}_2[m-1]) \\
&\quad + \sum_{m=1}^{n} H(Y_3[m]|W_2,\overline{Y}_3[m+1]) \\
&\quad - \sum_{m=1}^{n} H(Y_3[m]|W,W_2,\overline{Y}_3[m+1]) && (54) \\
&\leq n\epsilon_n + \sum_{m=1}^{n} H(Y_2[m]|\underline{Y}_2[m-1]) \\
&\quad - \sum_{m=1}^{n} H(Y_2[m]|W_2,\underline{Y}_2[m-1]) \\
&\quad + \sum_{m=1}^{n} H(Y_3[m]|W_2,\overline{Y}_3[m+1]) \\
&\quad - \sum_{m=1}^{n} H(Z_3[m]|W,W_2,\overline{Y}_3[m+1]) && (55) \\
&= n\epsilon_n + \sum_{m=1}^{n} H(Y_2[m]|\underline{Y}_2[m-1],X_2[m]) \\
&\quad - \sum_{m=1}^{n} H(Y_2[m]|W_2,\underline{Y}_2[m-1],X_2[m]) \\
&\quad + \sum_{m=1}^{n} H(Y_3[m]|W_2,\overline{Y}_3[m+1],X_2[m]) \\
&\quad - \sum_{m=1}^{n} H(Z_3[m]|W,W_2,\overline{Y}_3[m+1],X_2[m]) && (56) \\
&= n\epsilon_n + \sum_{m=1}^{n} I(W_2;Y_2[m]|\underline{Y}_2[m-1],X_2[m]) \\
&\quad + \sum_{m=1}^{n} I(W;Y_3[m]|W_2,\overline{Y}_3[m+1],X_2[m]) && (57)
\end{aligned}
$$

In (55) we used that $h(Y_3[m]) = h(c_{32}X_2[m] + \mathbf{c}_{31}^H\mathbf{X}_1[m] + Z_3[m]) \geq h(Z_3[m])$ (in a conditional version), and in (56) that $X_2[m]$ depends only on $\underline{Y}_2[m-1]$. We now define $U_2[m] = (W_2, \underline{Y}_2[m-1], \overline{Y}_3[m+1])$, and use Csiszar's identity as in [19] (2.5-2.8) to obtain (48).

We will prove the theorem for the synchronous case. The proof in the phase-fading case is a simpler case that we will omit. Using lemma 2 we get the low power limit of (48-49) as

$$\begin{aligned}
\mathsf{R} &\leq \lim_{B\to\infty} \frac{\operatorname{var}[\mathbf{c}_{31}^H\mathbf{X}_1(B)|U_2(B),X_2(B)]}{N_0} \\
&\quad + \lim_{B\to\infty} \frac{\operatorname{var}[\mathbf{c}_{21}^H\mathbf{X}_1(B)|X_2(B)]}{N_0} \\
&\quad - \lim_{B\to\infty} \frac{\operatorname{var}[\mathbf{c}_{21}^H\mathbf{X}_1(B)|U_2(B),X_2(B)]}{N_0} \qquad (58)\\
\mathsf{R} &\leq \lim_{B\to\infty} \frac{\operatorname{var}[\mathbf{c}_{31}^H\mathbf{X}_1(B)+c_{32}^*X_2(B)]}{N_0} \qquad (59)
\end{aligned}$$

Let $\mathbf{u}$ be a unit vector in the direction of $\lim_{B\to\infty}\operatorname{cov}[\mathbf{X}_1(B),X_2(B)]$, and define

$$\begin{aligned}
\beta\sqrt{P_1}\mathbf{u} &= \lim_{B\to\infty} \frac{\operatorname{cov}[\mathbf{X}_1(B),X_2(B)]}{N_0\sqrt{\operatorname{var}[X_2(B)]}} \qquad (60)\\
\mathbf{X} &= \lim_{B\to\infty} \frac{\operatorname{cov}[\mathbf{X}_1(B)]}{N_0} \qquad (61)\\
\mathbf{A} &= \lim_{B\to\infty} \frac{\operatorname{cov}[\mathbf{X}_1(B)|X_2(B),U_2(B)]}{N_0} \qquad (62)\\
\mathbf{B} &= \mathbf{X} - \beta^2P_1\mathbf{u}\mathbf{u}^H - \mathbf{A} \qquad (63)
\end{aligned}$$

Using Lemma 4 we then obtain the following outer bound to the low power rate

$$\begin{aligned}
\mathsf{R} &\leq \mathbf{c}_{31}^H\mathbf{B}\mathbf{c}_{31} + \mathbf{c}_{21}^H\mathbf{A}\mathbf{c}_{21} \qquad (64)\\
\mathsf{R} &\leq \mathbf{c}_{31}^H\left(\mathbf{A}+\mathbf{B}+\beta^2P_1\mathbf{u}\mathbf{u}^H\right)\mathbf{c}_{31} + |c_{32}|^2P_2 \\
&\quad + 2\Re\left\{\beta c_{32}\mathbf{c}_{31}^H\mathbf{u}\right\}\sqrt{P_1P_2} \qquad (65)
\end{aligned}$$

subject to

$$\begin{aligned}
\operatorname{tr}\mathbf{A} + \operatorname{tr}\mathbf{B} + \beta^2P_1 &\leq P_1 \qquad (66)\\
\mathbf{A},\mathbf{B} &\succeq 0 \qquad (67)\\
\beta &\leq 1 \qquad (68)
\end{aligned}$$

on the other hand, if $\mathbf{A}$, $\mathbf{B}$, and $\beta$ satisfy (66-68), then (64-65) constitute an upper bound on the rate.

It is clear that (64-65) is maximized for $\mathbf{A} = P_{21}\frac{\mathbf{c}_{21}\mathbf{c}_{21}^H}{\|\mathbf{c}_{21}\|^2}$ for some positive constant $P_{21}$. Now notice that if the angle between $\mathbf{c}_{21}$ and $\mathbf{c}_{31}$ is acute (if not, we can just use $-\mathbf{c}_{21}$), the bounds are maximized when the off-diagonal elements of $\mathbf{B}$ are maximized, i.e., if $\mathbf{B}$ has rank one. Thus, we can put $\mathbf{B} = P_{31}/B\mathbf{v}\mathbf{v}^H$, where $\mathbf{v}$ is a unit vector rotated an angle $\theta$ from $\mathbf{c}_{21}$ in the real plane spanned by $\mathbf{c}_{21}, \mathbf{c}_{31}$, as any component outside this plane will not contribute to the bounds. Finally, the bounds are maximized for $\mathbf{u} = \frac{\mathbf{c}_{31}}{\|\mathbf{c}_{31}\|}$. It is now a straightforward calculation to get the bounds (45) and (46).

For the achievable rate we split the message $W$ into two independent parts $W_d$ and $W_r$. The message $W_d$ is transmitted directly to the destination using power $P_{31}$ and a rate

$$\mathsf{R}_d = (|c_{311}|^2 + |c_{312}|^2)P_{31} \tag{69}$$

The message $W_r$ is transmitted through the relay using block Markov encoding with a rate

$$\begin{aligned}\mathsf{R}_r = & \min\left\{(|c_{211}|^2 + |c_{212}|^2)\cos^2(\alpha-\theta)P_{21},\right.\\ & (|c_{311}|^2 + |c_{312}|^2)\cos^2(\theta)P_{21}\\ & \left.+\left(\sqrt{P_{b1}(|c_{311}|^2 + |c_{312}|^2)} + |c_{32}|\sqrt{P_2}\right)^2\right\}\end{aligned} \tag{70}$$

Adding up these rates achieves the upper bound. ■

The proof of Theorem 1 uses the following Lemma

*Lemma 4:* For any random variables $\mathbf{X}$ and $Y$ with first and second order moments

$$\mathrm{cov}[\mathbf{X}|Y] \prec \mathrm{cov}[\mathbf{X}] - \frac{\mathrm{cov}[\mathbf{X},Y]\mathrm{cov}[\mathbf{X},Y]^H}{\mathrm{var}[Y]} \tag{71}$$

*Proof:* We can assume that $\mathbf{X}$ and $Y$ are zero mean. First, notice that

$$\mathrm{cov}[\mathbf{X}|Y] = E\left[\mathbf{X}\mathbf{X}^H\right] - E\left[E[\mathbf{X}|Y]E[\mathbf{X}|Y]^H\right] \tag{72}$$

Second, the Cauchy-Schwartz inequality gives

$$\begin{aligned}\left|E\left[\mathbf{v}^H\mathbf{X}Y\right]\right| &= \left|E\left[E\left[\mathbf{v}^H\mathbf{X}|Y\right]Y\right]\right|\\ &\leq \sqrt{E\left[|Y|^2\right]}\\ &\quad \sqrt{E\left[\mathbf{v}^H E[\mathbf{X}|Y]E[\mathbf{X}|Y]^H\mathbf{v}\right]}\end{aligned} \tag{73}$$

So,

$$\mathbf{v}^H E\left[\mathbf{X}Y^*\right] E\left[Y\mathbf{X}^H\right]\mathbf{v} \leq \mathrm{var}[Y]\mathbf{v}^H E\left[E[\mathbf{X}|Y]E[\mathbf{X}|Y]^H\right]\mathbf{v} \tag{74}$$

Which means $\mathrm{cov}\left[\mathbf{X},Y\right]\mathrm{cov}\left[\mathbf{X},Y\right]^H \prec \mathrm{var}[Y]E\left[E[\mathbf{X}|Y]E[\mathbf{X}|Y]^H\right]$. Inserting this gives (71). ■

We now turn to the relay channel in Fig. 1(b). From the two relays to the destination we have a MAC channel. As opposed to the usual MAC channel, we have messages that can have arbitrary dependency. The usual MAC outer bound can be generalized as follows, with the difference being that $X_2$ and $X_3$ can no longer be assumed independent

*Proposition 2:* The capacity region of of the multiple access channel with dependent messages is contained in the convex closure of all rates satisfying

$$R_{23} \leq I(X_2; Y_4|X_3) \tag{75}$$

$$R_{32} \leq I(X_3; Y_4|X_2) \tag{76}$$

$$R \leq I(X_3, X_3; Y_4) \tag{77}$$

for some joint distribution $p(x_2, x_3)$ that satisfies the power constraints.

The proof follows the usual MAC proof in [11, Theorem 14.3.3], just replacing $H(W_2)$ with $H(W_2|W_3)$ and $H(W_3)$ with $H(W_3|W_2)$.

Since the above bound, as the usual Gaussian MAC bound, is maximized by the Gaussian distribution, we get directly the bound in the low power regime

*Corollary 1:* The capacity region of the MAC channel in the low power regime in the synchronous case is contained in the convex closure of all all $(\mathsf{R}, \mathsf{R}_{23}, \mathsf{R}_{32})$ that satisfies

$$\mathsf{R}_{23} \leq |c_{42}|^2 P_2(1-\rho^2) \tag{78}$$

$$\mathsf{R}_{32} \leq |c_{43}|^2 P_3(1-\rho^2) \tag{79}$$

$$\mathsf{R} \leq |c_{42}|^2 P_2 + |c_{43}|^2 P_3 + 2\rho |c_{42}||c_{43}|\sqrt{P_2 P_3} \tag{80}$$

for some $\rho \in [0,1]$ in the synchronous case.

In the phase-fading case the rates satisfy

$$\mathsf{R}_{23} \leq |c_{42}|^2 P_2 \tag{81}$$

$$\mathsf{R}_{32} \leq |c_{43}|^2 P_3 \tag{82}$$

$$\mathsf{R} \leq |c_{42}|^2 P_2 + |c_{43}|^2 P_3 \tag{83}$$

*Theorem 2:* In the phase fading case, the deterministic capacity of the relay channel in Fig. 1 is given by transmitting a common message to the two relays in addition to two private messages.

*Proof:* We will prove that upper bound for the broadcast part of the channel can be achieved with a common/private message transmission scheme. Since this is clearly also true for the MAC part, this will be sufficient to prove the Theorem. We will prove the theorem for the case when antenna 1 and antenna 2 of node 1 have separate power constraints, which we will denote $P_1$ and $P_2$. The result then clearly applies to the case when there is a sum power constraint, but it also applies to the case when the two antennas are actually on separate nodes.

For the broadcast part of the channel, the achievable rate by common/private message transmission is given by

$$\begin{aligned}
\mathsf{R}_c &= \min\left\{|c_{211}|^2 P_{1c} + |c_{212}|^2 P_{2c}, |c_{311}|^2 P_{1c} + |c_{312}|^2 P_{2c}\right\} \\
\mathsf{R}_2 &= \mathsf{R}_c + |c_{211}|^2 P_{12} + |c_{212}|^2 P_{22} \\
\mathsf{R}_3 &= \mathsf{R}_c + |c_{311}|^2 P_{13} + |c_{312}|^2 P_{23} \\
\mathsf{R} &= \mathsf{R}_c + |c_{211}|^2 P_{12} + |c_{212}|^2 P_{22} + |c_{311}|^2 P_{13} + |c_{312}|^2 P_{23}
\end{aligned}$$

with the constraints

$$\begin{aligned}
P_{1c} + P_{21} + P_{31} &\leq P_1 \\
P_{2c} + P_{22} + P_{32} &\leq P_2 \qquad (84)
\end{aligned}$$

For the upper bound we apply Lemma 2 to the bounds of Proposition 1

$$\begin{aligned}
\mathsf{R}_2 &\leq |c_{211}|^2 \lim_{B\to\infty} \frac{\mathrm{var}[X_{11}(B)] - \mathrm{var}[X_{11}(B)|U_2(B)]}{N_0} \\
&\quad + |c_{212}|^2 \lim_{B\to\infty} \frac{\mathrm{var}[X_{12}(B)] - \mathrm{var}[X_{12}(B)|U_2(B)]}{N_0} &(85)\\
\mathsf{R}_3 &\leq |c_{311}|^2 \lim_{B\to\infty} \frac{\mathrm{var}[X_{11}(B)] - \mathrm{var}[X_{11}(B)|U_3(B)]}{N_0} \\
&\quad + |c_{312}|^2 \lim_{B\to\infty} \frac{\mathrm{var}[X_{12}(B)] - \mathrm{var}[X_{12}(B)|U_3(B)]}{N_0} &(86)\\
\mathsf{R} &\leq |c_{311}|^2 \lim_{B\to\infty} \frac{\mathrm{var}[X_{11}(B)] - \mathrm{var}[X_{11}(B)|U_3(B)]}{N_0} \\
&\quad + |c_{211}|^2 \lim_{B\to\infty} \frac{\mathrm{var}[X_{11}(B)|U_3(B)]}{N_0} \\
&\quad + |c_{312}|^2 \lim_{B\to\infty} \frac{\mathrm{var}[X_{12}(B)] - \mathrm{var}[X_{12}(B)|U_3(B)]}{N_0} \\
&\quad + |c_{212}|^2 \lim_{B\to\infty} \frac{\mathrm{var}[X_{12}(B)|U_3(B)]}{N_0} &(87)\\
\mathsf{R} &\leq |c_{211}|^2 \lim_{B\to\infty} \frac{\mathrm{var}[X_{11}(B)] - \mathrm{var}[X_{11}(B)|U_2(B)]}{N_0} \\
&\quad + |c_{311}|^2 \lim_{B\to\infty} \frac{\mathrm{var}[X_{11}(B)|U_2(B)]}{N_0} \\
&\quad + |c_{212}|^2 \lim_{B\to\infty} \frac{\mathrm{var}[X_{12}(B)] - \mathrm{var}[X_{12}(B)|U_2(B)]}{N_0} \\
&\quad + |c_{312}|^2 \lim_{B\to\infty} \frac{\mathrm{var}[X_{12}(B)|U_2(B)]}{N_0} &(88)
\end{aligned}$$

Define

$$\begin{aligned}
P_{12} &= \lim_{B\to\infty} \frac{\mathrm{var}[X_{11}(B)|U_3(B)]}{N_0} &(89)\\
P_{22} &= \lim_{B\to\infty} \frac{\mathrm{var}[X_{12}(B)|U_3(B)]}{N_0} &(90)\\
P_{13} &= \lim_{B\to\infty} \frac{\mathrm{var}[X_{11}(B)|U_2(B)]}{N_0} &(91)\\
P_{23} &= \lim_{B\to\infty} \frac{\mathrm{var}[X_{12}(B)|U_2(B)]}{N_0} &(92)\\
P_{1c} &= \lim_{B\to\infty} \frac{\mathrm{var}[X_{11}(B)] - \mathrm{var}[X_{11}(B)|U_3(B)]}{N_0} \\
&\quad - \lim_{B\to\infty} \frac{\mathrm{var}[X_{11}(B)|U_2(B)]}{N_0} &(93)\\
P_{2c} &= \lim_{B\to\infty} \frac{\mathrm{var}[X_{12}(B)] - \mathrm{var}[X_{12}(B)|U_3(B)]}{N_0} \\
&\quad - \lim_{B\to\infty} \frac{\mathrm{var}[X_{12}(B)|U_2(B)]}{N_0} &(94)
\end{aligned}$$

Clearly $P_{ij} \geq 0$, so that we can think of them as powers. Notice that we cannot assume $P_{ic} \geq 0$. However, we

have the constraints

$$P_{1c} + P_{12} + P_{13} \leq P_1 \quad (95)$$

$$P_{2c} + P_{22} + P_{23} \leq P_2 \quad (96)$$

$$P_{1c} + P_{12} \geq 0 \quad (97)$$

$$P_{1c} + P_{13} \geq 0 \quad (98)$$

$$P_{2c} + P_{22} \geq 0 \quad (99)$$

$$P_{2c} + P_{23} \geq 0 \quad (100)$$

With this we can write

$$\begin{aligned} \mathsf{R}_3 &\leq |c_{211}|^2 P_{1c} + |c_{212}|^2 P_{2c} \\ &+ |c_{211}|^2 P_{12} + |c_{212}|^2 P_{22} \quad (101) \\ \mathsf{R}_4 &\leq |c_{311}|^2 P_{1c} + |c_{312}|^2 P_{2c} \\ &+ |c_{311}|^2 P_{13} + |c_{312}|^2 P_{23} \quad (102) \\ \mathsf{R} &\leq |c_{211}|^2 P_{1c} + |c_{212}|^2 P_{2c} + |c_{211}|^2 P_{12} \\ &+ |c_{212}|^2 P_{22} + |c_{311}|^2 P_{13} + |c_{312}|^2 P_{23} \quad (103) \\ \mathsf{R} &\leq |c_{311}|^2 P_{1c} + |c_{312}|^2 P_{2c} + |c_{211}|^2 P_{12} \\ &+ |c_{212}|^2 P_{22} + |c_{311}|^2 P_{13} + |c_{312}|^2 P_{23} \quad (104) \end{aligned}$$

We will show that the upper bound can always be achieved by a common/private message solution. First consider the case $\{|c_{211}|, |c_{212}|\} \leq \{|c_{311}|, |c_{312}|\}$. The optimum solution has $P_{12} = P_{22} = 0$. Namely, putting $P_{1c} \to P_{1c} + P_{12}$ and $P_{2c} \to P_{2c} + P_{22}$ will not decrease any rate bounds, while the power bounds are still satisfied. Notice that we can now assume $P_{1c} \geq 0$, $P_{2c} \geq 0$. So, we end up with

$$\begin{aligned} \mathsf{R}_3 &\leq |c_{211}|^2 P_{1c} + |c_{212}|^2 P_{2c} \quad (105) \\ \mathsf{R}_4 &\leq |c_{311}|^2 P_{1c} + |c_{312}|^2 P_{2c} \\ &+ |c_{311}|^2 P_{13} + |c_{312}|^2 P_{23} \quad (106) \\ \mathsf{R} &\leq |c_{311}|^2 P_{1c} + |c_{312}|^2 P_{2c} \\ &+ |c_{311}|^2 P_{13} + |c_{312}|^2 P_{23} \quad (107) \\ \mathsf{R} &\leq |c_{211}|^2 P_{1c} + |c_{212}|^2 P_{2c} \\ &+ |c_{311}|^2 P_{13} + |c_{312}|^2 P_{23} \quad (108) \end{aligned}$$

or

$$\begin{aligned} \mathsf{R}_3 &\le |c_{211}|^2 P_{1c} + |c_{212}|^2 P_{2c} && (109) \\ \mathsf{R}_4 &\le |c_{211}|^2 P_{1c} + |c_{212}|^2 P_{2c} \\ &\quad + |c_{311}|^2 P_{13} + |c_{312}|^2 P_{23} && (110) \\ \mathsf{R} &\le |c_{211}|^2 P_{1c} + |c_{212}|^2 P_{2c} \\ &\quad + |c_{311}|^2 P_{13} + |c_{312}|^2 P_{23} && (111) \end{aligned}$$

This can be achieved by transmitting a common message understood by both nodes, and a private message to node 4. The symmetric case is similar.

Next consider the case $|c_{211}| \le |c_{311}|, |c_{312}| \le |c_{212}|$, with strict inequality in at least one of the inequalities. Then a solution with $P_{12} = P_{23} = 0$ is optimum, which can be seen by putting $P_{1c} \to P_{1c} + P_{12}$ and $P_{2c} \to P_{2c} + P_{23}$. Again we can then assume $P_{1c} \ge 0$, $P_{2c} \ge 0$. Then

$$\begin{aligned} \mathsf{R}_3 &\le |c_{211}|^2 P_{1c} + |c_{212}|^2 P_{2c} + |c_{212}|^2 P_{22} && (112) \\ \mathsf{R}_4 &\le |c_{311}|^2 P_{1c} + |c_{312}|^2 P_{2c} + |c_{311}|^2 P_{13} && (113) \\ \mathsf{R} &\le |c_{311}|^2 P_{1c} + |c_{312}|^2 P_{2c} \\ &\quad + |c_{311}|^2 P_{13} + |c_{212}|^2 P_{22} && (114) \\ \mathsf{R} &\le |c_{211}|^2 P_{1c} + |c_{212}|^2 P_{2c} \\ &\quad + |c_{311}|^2 P_{13} + |c_{212}|^2 P_{22} && (115) \end{aligned}$$

We will argue that we can always obtain an optimum solution with the right hand sides of (114) and (115) equal. Assume the right hand side of (114) is smaller than that of (115). We can decrease (115) by putting $P_{13} \to P_{13} - \delta$, $P_{1c} \to P_{1c} + \delta$. Either the bounds become equal, or we end up with $P_{13} = 0$, so

$$\begin{aligned} \mathsf{R}_3 &\le |c_{211}|^2 P_{1c} + |c_{212}|^2 P_{2c} + |c_{212}|^2 P_{22} && (116) \\ \mathsf{R}_4 &\le |c_{311}|^2 P_{1c} + |c_{312}|^2 P_{2c} && (117) \\ \mathsf{R} &\le |c_{311}|^2 P_{1c} + |c_{312}|^2 P_{2c} + |c_{212}|^2 P_{22} && (118) \\ \mathsf{R} &\le |c_{211}|^2 P_{1c} + |c_{212}|^2 P_{2c} + |c_{212}|^2 P_{22} && (119) \end{aligned}$$

But this can be written as

$$\begin{aligned} \mathsf{R}_3 &\le |c_{311}|^2 P_{1c} + |c_{312}|^2 P_{2c} + |c_{212}|^2 P_{22} && (120) \\ \mathsf{R}_4 &\le |c_{311}|^2 P_{1c} + |c_{312}|^2 P_{2c} && (121) \\ \mathsf{R} &\le |c_{311}|^2 P_{1c} + |c_{312}|^2 P_{2c} + |c_{212}|^2 P_{22} && (122) \end{aligned}$$

which can be achieved by a common message and a private message to node 3.

On the other hand, suppose the right hand side of (114) is larger than that of (115). Then we can decrease (114)

by by putting $P_{13} \to P_{22} - \delta$, $P_{2c} \to P_{2c} + \delta$. If the two bounds don't become equal we end up with

$$\mathsf{R}_3 \leq |c_{211}|^2 P_{1c} + |c_{212}|^2 P_{2c} \tag{123}$$

$$\mathsf{R}_4 \leq |c_{311}|^2 P_{1c} + |c_{312}|^2 P_{2c} + |c_{311}|^2 P_{13} \tag{124}$$

$$\mathsf{R} \leq |c_{311}|^2 P_{1c} + |c_{312}|^2 P_{2c} + |c_{311}|^2 P_{13} \tag{125}$$

$$\mathsf{R} \leq |c_{211}|^2 P_{1c} + |c_{212}|^2 P_{2c} + |c_{311}|^2 P_{13} \tag{126}$$

which, as above, can be achieved by common/private messaging. ■

*Theorem 3:* Consider the relay channel in Fig. 1(b) for the synchronous case. Assume that

$$\min\{\|\mathbf{c}_{21}\|^2, \|\mathbf{c}_{31}\|^2\} \leq |\mathbf{c}_{21}^H \mathbf{c}_{31}|. \tag{127}$$

The deterministic capacity is then given by transmitting a common message to the relays and a private message to one of the relays.

*Proof:* We will prove that the upper bound for the broadcast cut set can be achieved by a common/private message solution. Without loss of generality we can consider the case $\|\mathbf{c}_{21}\| \geq \|\mathbf{c}_{31}\|$. According to Lemma 1 and 2 we have

$$\begin{aligned}
\mathsf{R}_2 \leq\ & \lim_{B\to\infty} \frac{B\mathrm{var}[\mathbf{c}_{21}^H \mathbf{X}(B)]}{N_0} \\
& - \lim_{B\to\infty} \frac{B\mathrm{var}[\mathbf{c}_{21}^H \mathbf{X}(B)|U_2(B)]}{N_0}
\end{aligned} \tag{128}$$

$$\begin{aligned}
\mathsf{R}_3 \leq\ & \lim_{B\to\infty} \frac{B\mathrm{var}[\mathbf{c}_{31}^H \mathbf{X}(B)]}{N_0} \\
& - \lim_{B\to\infty} \frac{B\mathrm{var}[\mathbf{c}_{31}^H \mathbf{X}(B)|U_3(B)]}{N_0}
\end{aligned} \tag{129}$$

and

$$\begin{aligned}
\mathsf{R} \leq\ & \lim_{B\to\infty} \frac{B\mathrm{var}[\mathbf{c}_{21}^H \mathbf{X}(B)|U_3(B)]}{N_0} \\
& + \lim_{B\to\infty} \frac{B\mathrm{var}[\mathbf{c}_{31}^H \mathbf{X}(B)]}{N_0} \\
& - \lim_{B\to\infty} \frac{B\mathrm{var}[\mathbf{c}_{31}^H \mathbf{X}(B)|U_3(B)]}{N_0}
\end{aligned} \tag{130}$$

$$\begin{aligned}
\mathsf{R} \leq\ & \lim_{B\to\infty} \frac{B\mathrm{var}[\mathbf{c}_{31}^H \mathbf{X}(B)|U_2(B)]}{N_0} \\
& + \lim_{B\to\infty} \frac{B\mathrm{var}[\mathbf{c}_{21}^H \mathbf{X}(B)]}{N_0} \\
& - \lim_{B\to\infty} \frac{B\mathrm{var}[\mathbf{c}_{21}^H \mathbf{X}(B)|U_2(B)]}{N_0}
\end{aligned} \tag{131}$$

As in the proof of Theorem 2 we can upper bound this by

$$\begin{aligned} \mathsf{R}_2 &\leq \mathbf{c}_{21}^H (\mathbf{X}-\mathbf{A})\mathbf{c}_{21} \\ \mathsf{R}_3 &\leq \mathbf{c}_{31}^H (\mathbf{X}-\mathbf{B})\mathbf{c}_{31} \\ \mathsf{R} &\leq \mathbf{c}_{31}^H (\mathbf{X}-\mathbf{B})\mathbf{c}_{31} + \mathbf{c}_{21}^H\mathbf{B}\mathbf{c}_{21} \\ \mathsf{R} &\leq \mathbf{c}_{21}^H (\mathbf{X}-\mathbf{A})\mathbf{c}_{21} + \mathbf{c}_{31}^H\mathbf{A}\mathbf{c}_{31} \end{aligned} \tag{132}$$

with the conditions

$$\mathrm{tr}\mathbf{X} \leq P \tag{133}$$

$$\mathbf{X} \succ \mathbf{A},\mathbf{B} \tag{134}$$

$$\mathbf{A},\mathbf{B} \succ \mathbf{0} \tag{135}$$

Notice that the condition $\|\mathbf{c}_{21}\|^2 \geq \left|\mathbf{c}_{21}^H\mathbf{c}_{31}\right| \geq \|\mathbf{c}_{31}\|^2$ ensures that $\left|\mathbf{c}_{21}^H\mathbf{v}\right| \geq \left|\mathbf{c}_{31}^H\mathbf{v}\right|$ for any vector $\mathbf{v}$. Consequently $\mathbf{c}_{21}^H\mathbf{A}\mathbf{c}_{21} \geq \mathbf{c}_{31}^H\mathbf{A}\mathbf{c}_{31}$, and we can put $\mathbf{A}=\mathbf{0}$ without decreasing any bounds. The bounds then are

$$\mathsf{R}_2 \leq \mathbf{c}_{21}^H\mathbf{X}\mathbf{c}_{21} \tag{136}$$

$$\mathsf{R}_3 \leq \mathbf{c}_{31}^H (\mathbf{X}-\mathbf{B})\mathbf{c}_{31} \tag{137}$$

$$\mathsf{R} \leq \mathbf{c}_{31}^H (\mathbf{X}-\mathbf{B})\mathbf{c}_{31} + \mathbf{c}_{21}^H\mathbf{B}\mathbf{c}_{21} \tag{138}$$

$$\mathsf{R} \leq \mathbf{c}_{21}^H\mathbf{X}\mathbf{c}_{21} \tag{139}$$

Here the right hand side of (138) is clearly larger than the right hand side of (139), so we can rewrite as

$$\mathsf{R}_2 \leq \mathbf{c}_{31}^H\mathbf{C}\mathbf{c}_{31} + \mathbf{c}_{21}^H\mathbf{B}\mathbf{c}_{21} \tag{140}$$

$$\mathsf{R}_3 \leq \mathbf{c}_{31}^H\mathbf{C}\mathbf{c}_{31} \tag{141}$$

$$\mathbf{X} = \mathbf{C}+\mathbf{B} \tag{142}$$

It is clear that this is optimized for $\mathbf{B}=\alpha_2\mathbf{c}_{21}\mathbf{c}_{21}^H$ and $\mathbf{C}=\alpha_3\mathbf{c}_{31}\mathbf{c}_{31}^H$, and that this can be achieve by transmitting a private message in the direction of $\mathbf{c}_{21}$ and a common message in the direction of $\mathbf{c}_{31}$. ■

On the other hand, it is shown in the Appendix that Theorem 3 is not necessarily true if the condition (127) is not satisfied.

## V. Concluding remarks

In this paper the deterministic capacity has been found for two MIMO relay networks in the low power regime. The methodology certainly can be extended to other networks. For example, it is not difficult to see that the results for the networks in Figure 1 can be generalized to the case when nodes have more antennas. Also, some links and extra relays can be carefully added. The paper [7] shows how deterministic capacity can be generalized to a two node cooperative MAC channel. That paper also shows that the functions of messages are not always simply sub-messages (already [8] shows that more general function of messages are needed).

On the other hand, it is very difficult to find the deterministic capacity of larger networks in general with the present methodology. For example, if a link between the relays in Figure 1(b) or between source and destination is added, or if more relays are added, there is not straightforward way of generalizing the results presented here.

There are two issues that make generalization difficult. First, it is very difficult to find single letter bounds.

For example, there is no direct way to generalize Nair and El Gamal's bound [19]or Marton's bound [20] for the broadcast channel to more than two nodes. Non-single letter bounds for general networks can easily be found by using Fano's inequality, $R_i \leq I(W_i; Y_i)$ and $R_{ij} \leq I(W_i; Y_i|W_j)$. But there seems to be no systematic way of generating single letter bounds from these for larger networks.

Secondly, even when single letter bounds exist, they may not be tight, as we have seen for the network 1(b) in the synchronous case. This could of course just be because Nair and El Gamal's bound is not the best possible single letter bound. However, this *appears* to be a more fundamental issue. Single letter bounds essentially show that iid (independent, identically distributed) input is optimum. Instead take non-single letter bounds such as $R_i \leq I(W_i; Y_i)$ and $R_{ij} \leq I(W_i; Y_i|W_j)$ and *assume* that iid input is optimum. Then the counter example in the Appendix still works. This indicates that either there is better transmission scheme than common/private messages, or that Fano-type bounds are not tight.

## APPENDIX

In this appendix we will show that the upper bound of Theorem 1 is not necessarily achievable with common/private message transmission in the synchronous case. Specifically we will prove that the bound (132) is not achievable by common/private message transmission. Strictly speaking it does not prove that Theorem 1 is not achievable, as it is not proven that any set of positive definite matrices $(\mathbf{X},\mathbf{A},\mathbf{B})$ (with $\mathbf{X}\succ\mathbf{A},\mathbf{B}$) is a valid set of covariance matrices. However, it indicates that the proof technique does not work in general for the synchronous case.

We put $\|\mathbf{c}_{21}\|^2=\|\mathbf{c}_{31}\|^2=1$ and $\angle(\mathbf{c}_{21},\mathbf{c}_{31})=\alpha=0.4$. We consider the real subspace spanned by $\mathbf{c}_{21},\mathbf{c}_{31}$, and in this define an orthonormal basis by $\{\mathbf{c}_{21},\mathbf{c}_{21}^{\perp}\}$. Then

$$\mathbf{c}_{31} = \begin{bmatrix}\cos\alpha\\ \sin\alpha\end{bmatrix} \quad (143)$$

$$= \begin{bmatrix}0.9211\\ 0.3894\end{bmatrix} \quad (144)$$

We now define

$$\mathbf{v} = \begin{bmatrix}\cos 0.208\\ \sin 0.208\end{bmatrix} = \begin{bmatrix}0.9784\\ 0.2065\end{bmatrix} \quad (145)$$

$$\mathbf{X}-\mathbf{B} = \mathbf{v}\mathbf{v}^H \quad (146)$$

$$\mathbf{B} = \mathbf{c}_{21}\mathbf{c}_{21}^H \quad (147)$$

$$\mathbf{A} = 0.05\mathbf{c}_{31}\mathbf{c}_{31}^H \quad (148)$$

$$\mathbf{X} = \mathbf{B}+(\mathbf{X}-\mathbf{B}) = \mathbf{v}\mathbf{v}^H+\mathbf{c}_{21}\mathbf{c}_{21}^H \quad (149)$$

$$\mathbf{X}-\mathbf{A} = \mathbf{v}\mathbf{v}^H+\mathbf{c}_{21}\mathbf{c}_{21}^H-0.05\mathbf{c}_{31}\mathbf{c}_{31}^H \quad (150)$$

$$= \begin{bmatrix}1.9149 & 0.1841\\ 0.1841 & 0.03506\end{bmatrix} \quad (151)$$

The eigenvalues of $\mathbf{X}-\mathbf{A}$ are $(0.01720, 1.9328)$, so $\mathbf{X}-\mathbf{A}$ is positive definite. We now have

$$\begin{aligned}
\mathsf{R}_2 &= \mathbf{c}_{21}^H(\mathbf{X}-\mathbf{A})\mathbf{c}_{21}\\
&= \cos^2(0.208)+1-0.05\cos^2\alpha \;=\; 1.9149\\
\mathsf{R}_3 &= \mathbf{c}_{31}^H(\mathbf{X}-\mathbf{B})\mathbf{c}_{31}\\
&= \cos^2(\alpha-0.208) \;=\; 0.9636\\
\mathsf{R}_a &= \mathbf{c}_{31}^H(\mathbf{X}-\mathbf{B})\mathbf{c}_{31}+\mathbf{c}_{21}^H\mathbf{B}\mathbf{c}_{21}\\
&= 0.9636+1 \;=\; 1.9636\\
\mathsf{R}_b &= \mathbf{c}_{21}^H(\mathbf{X}-\mathbf{A})\mathbf{c}_{21}+\mathbf{c}_{31}^H\mathbf{A}\mathbf{c}_{31}\\
&= 1.9149+0.05 \;=\; 1.9649 \quad (152)
\end{aligned}$$

Consider the achievable rate by a common/private messaging scheme in the Gaussian channel. We transmit the common message along a unit vector $\mathbf{u}$ and beamform the private messages to their respective destinations. This scheme achieves the following rates

$$\mathsf{R}_0 \leq \min\{\mathbf{c}_{21}^H\mathbf{u}\mathbf{u}\mathbf{c}_{21}, \mathbf{c}_{31}^H\mathbf{u}\mathbf{u}\mathbf{c}_{31}\}P_0 \quad (153)$$

$$\mathsf{R}_2 \leq \|\mathbf{c}_{21}\|^2 P_2 + R_0 \quad (154)$$

$$\mathsf{R}_3 \leq \|\mathbf{c}_{31}\|^2 P_3 + R_0 \quad (155)$$

$$\mathsf{R} \leq \|\mathbf{c}_{21}\|^2 P_2 + \|\mathbf{c}_{31}\|^2 P_3 + R_0 \quad (156)$$

subject to the constraint $P_0 + P_2 + P_3 \leq P$. Put

$$c_0^2 = \max_{\|\mathbf{u}\|=1} \min\{\mathbf{c}_{21}^H\mathbf{u}\mathbf{u}\mathbf{c}_{21}, \mathbf{c}_{31}^H\mathbf{u}\mathbf{u}\mathbf{c}_{31}\} \quad (157)$$

In this case it's easy to see that

$$c_0^2 = \cos^2(\alpha/2) = 0.9605 \quad (158)$$

Now consider the problem of given a rate triple $(\mathsf{R}_2, \mathsf{R}_3, \mathsf{R})$ minimizing the total power $P$. We have to solve

$$\min\left\{\frac{\mathsf{R}_0}{c_0} + \frac{\mathsf{R}_2'}{\|\mathbf{c}_2\|^2} + \frac{\mathsf{R}_3'}{\|\mathbf{c}_3\|^2}\right\} \quad (159)$$

subject to

$$\mathsf{R}_2' + \mathsf{R}_0 \geq \mathsf{R}_2 \quad (160)$$

$$\mathsf{R}_3' + \mathsf{R}_0 \geq \mathsf{R}_3 \quad (161)$$

$$\mathsf{R}_2' + \mathsf{R}_3' + \mathsf{R}_0 \geq \mathsf{R} \quad (162)$$

It's easy to see that the optimum solution is

$$P = \frac{\mathsf{R}_2 + \mathsf{R}_3 - \mathsf{R}}{c_0} + \frac{\mathsf{R} - \mathsf{R}_3}{\|\mathbf{c}_2\|^2} + \frac{\mathsf{R} - \mathsf{R}_2}{\|\mathbf{c}_3\|^2} \quad (163)$$

Inserting $(\mathsf{R}_2, \mathsf{R}_2, \mathsf{R}) = (1.9149, 0.9636, 1.9636)$ from (152) we get

$$\begin{aligned} P &= \frac{\mathsf{R}_2 + \mathsf{R}_3 - \mathsf{R}}{c_0^2} + \mathsf{R} - \mathsf{R}_2 + \mathsf{R} - \mathsf{R}_3 \\ &= 2.0011 > \mathrm{tr}\mathbf{X} = 2 \quad (164) \end{aligned}$$

which shows that the solution (152) is not achievable by common/private message transmission.